\documentclass[twocolumn,superscriptaddress,10pt,aps,showpacs,reprint]{revtex4-1}

\usepackage{amsmath}
\usepackage{amssymb}
\usepackage{microtype}
\usepackage{graphicx}

\usepackage{color}

\begin{document}

\title{Balancing building and maintenance costs in growing transport networks}

\author{Arianna Bottinelli}
\affiliation{Mathematics Department, Uppsala University, L\"agerhyddsv\"agen 1, Uppsala 75106, Sweden.}

\author{R\'emi Louf}
\affiliation{Centre for Advanced Spatial Analysis, University College London, 90
Tottenham Court Road W1T4TJ London, United Kingdom.}

\author{Marco Gherardi}
\affiliation{Sorbonne Universit\'es, UPMC Univ Paris 06, UMR 7238,
  Computational and Quantitative Biology, 15 rue de l'\'{E}cole de
  M\'{e}decine Paris, France.}
\affiliation{Dipartimento di Fisica, Universit\`a degli Studi di Milano, via Celoria 16, 20133 Milano, Italy.}

\begin{abstract}

\noindent The costs associated to the length of links impose unavoidable constraints to the growth of
    natural and artificial transport networks. 
When future network developments can not be predicted,
  building and maintenance costs require competing minimization mechanisms, and can not be optimized simultaneously. 
 Hereby, we study the interplay of building and maintenance costs 
  and its impact on the growth of transportation networks
  through a non-equilibrium model of network growth.
   We show cost balance is a sufficient ingredient for the emergence of 
  tradeoffs between the network's total length and transport efficiency,
   of optimal strategies of construction, 
   and of power-law temporal correlations in the growth history of the network.
    Analysis of empirical
    ant transport networks in the framework of this model
    suggests different ant species may adopt similar optimization strategies.
\end{abstract}


\maketitle

From roads, railways and power grids, to ant trails, leaf veins and blood vessels, 
transportation structures support the functions necessary to many natural and man-made systems~\cite{katifori2016, tero2010rules, perna2012, gastner2006shape, buhl2006topological, Gorman2004,vessels1,bebber2007biological,buhl2009shape}. 
Transport systems are typically represented as spatial networks, where nodes are distinct locations --- such as cities or ant nests --- and links are physical connections between these locations --- such as roads or trails~\cite{xie2009modeling}.
As transport networks are embedded in a metric space, the length of links is used to quantify the cost of building and maintaining the connections~\cite{Barthelemy2011}.
These costs pose an unavoidable constraint
to transport networks, which is intrinsically tied to their spatial nature.
Together with the need for efficient transportation and for fault tolerance, costs
affect the growth and the topology of transport networks, having profound impact on the systems that rely on them~\cite{Barthelemy2011}. 

A great deal of theoretical and empirical research in physics, quantitative geography, and transport engineering has been devoted to understand how diverse constraints influence the evolution of natural and man-made transport networks, and to identify minimal ingredients underlying the emergence of complex topologies~\cite{GastnerNewman:2006,BanavarMaritan:1999, MaritanColaiori:1996,Guimera:2002,Colizza:2004,Bottinelli2015}.
The effects of competing design criteria have been explored, such as average shortest path versus link density~\cite{FerreriCancho:2003} (or total length \cite{Brede:2010}), and total length versus synchronizability~\cite{Brede:2010b} or centrality~\cite{BarthelemyFlammini:2006}.
Other models balance the length of newly added links with the gain in centrality~\cite{fabrikant2002heuristically}, or efficiency~\cite{gastner2006shape}, or analyze the costs and benefits entailed by their creation~\cite{louf2013emergence}. 
However, most of the existing models assume that (i) the network is static and/or constituted by a pre-fixed and known set of nodes, (ii) it is either planned by a central authority, or the result of a completely self-organized process, and (iii) the length of a link is a proxy for both the costs of building \emph{and} maintaining it~\cite{Barthelemy2011}. Therefore, they neglect that (i) transport systems are typically built iteratively, often lacking information about future developments, as these may be beyond the time horizon of planners~\cite{xie2009modeling,Barthelemy2011};
(ii) due to such dynamic evolution, in long-lived infrastructures global planning has to compromise with local constraints and competing interests~\cite{xie2009modeling}, and to alternate with local optimization processes;
(iii) building costs and maintenance costs act on different time scales, constituting unavoidable competing constraints that cannot be optimized simultaneously.

These three aspects are strongly related.
In a static scenario,
the network of minimum length spanning a fixed set of nodes (the \emph{minimum spanning tree}, MST)
minimizes both maintenance and building costs~\cite{Barthelemy2011}.
In a dynamic setting, instead, when future node additions are not known in advance, or when the task of building links
is partially delegated to local entities, these costs can not be minimized simultaneously.
On one side, building cost is minimized by iterating the local rule of 
``linking each new node to the closest node in the network''.
However, the obtained structure (called \emph{dynamical minimum spanning tree}, dMST~\cite{fabrikant2002heuristically})
does not minimize the total length of the network~\cite{Bottinelli2015}, thus attaining a sub-optimal maintenance cost.
On the other side, globally rearranging the network to a MST every time a node is added
does minimize the total length, but it requires to destroy old links and rebuild new ones, increasing the building cost.
Moreover, maintenance costs must be sustained until links are abandoned or destroyed~\cite{1980transport, taplin2005cost}, 
constraining the network on a longer time scale.

\begin{figure}
\centering
\includegraphics{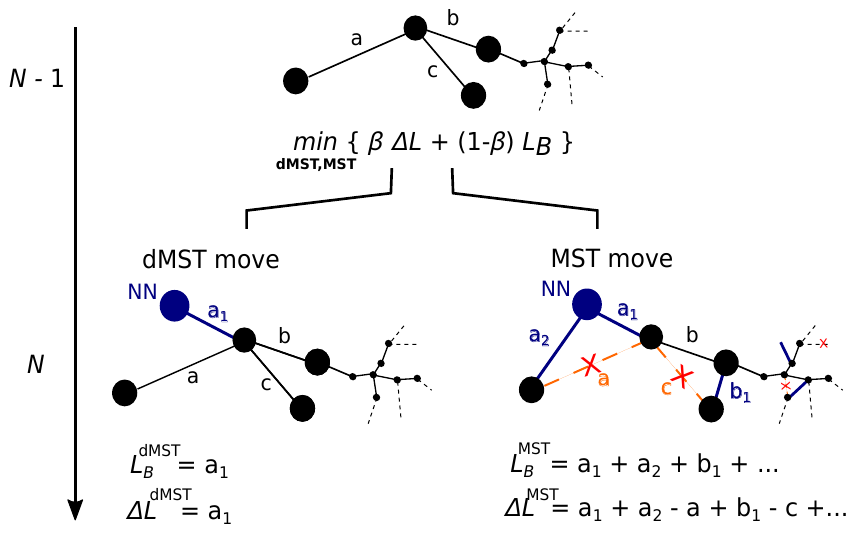}
\caption{(Color online) At each time step, the model grows a network by adding a new node (NN) at a random position. Depending on what move minimizes the linear combination of built length $L_B$ and length variation $\Delta L$, NN is either locally connected to the closest node in the network (dMST move) or the network is globally rewired to minimize the total length (MST move). In both cases only one link is added.}
\label{fig:1}
\end{figure}

In this paper, we address these open issues by formulating an out-of-equilibrium model for the growth of transport networks in the context where the position of new nodes can not be predicted. 
By combining the two pure optimization strategies (global, centrally planned MST, and local, decentralized dMST), the model explores the antagonism between the constraints associated with building and maintenance costs.

\textit{Model ---}
Our model grows spatial networks starting from a single node and adding one node and one link at a time, so that resulting networks are trees (see Fig.~\ref{fig:1}).
(In real transport networks, fault tolerance is achieved through the presence of cycles \cite{louzada2012generating,Barabasi2000}. 
Here we restrict to trees for simplicity.)
Nodes appear with the flat measure on the unit square.
When a new node at position $x_N$ is added to the existing nodes having positions $\{x_0,\ldots,x_{N-1}\}$,
either it is linked to the closest node (``dMST move''),
or a number of links are destroyed and rebuilt in order to obtain the (unique)
MST spanning all nodes at positions $\{x_0,\ldots,x_N\}$ (``MST move''), such that the functional
\begin{equation}
H(\beta, N) = \beta \Delta L(N) + (1- \beta) L_\mathrm{B}(N)
\end{equation}
is minimum. 
$L_\mathrm{B}$ is the length that needs to be built,
and $\Delta L$ is the variation in the total length of the network
(these are not equal, as $\Delta L$ includes negative contributions
from the deleted links).
To elaborate, 
both $H^{\mathrm{MST}}$ and $H^{\mathrm{dMST}}$ are computed every time a node is added, then
the MST move is performed if $H^{\mathrm{MST}} < H^{\mathrm{dMST}}$, and the dMST move otherwise.
The ``strategy'' $\beta$ is the only parameter of the model, taking values in $[0,1]$.
Setting $\beta=0$ prioritizes the minimization of $L_\mathrm{B}$ (as expected if building costs are dominant), 
and the network grows only by local dMST moves.
Conversely, $\beta=1$ minimizes $\Delta L$ (maintenance costs dominate), 
and the network is globally rewired to a MST at each step.
When the two costs are comparable, intermediate values of $\beta$ account for both global and local length minimization and the model can alternate between MST and dMST moves.
It is useful to
express the growth condition $H^{\mathrm{MST}}\gtrless H^{\mathrm{dMST}}$ in terms of the sum of the lengths of newly built and newly destroyed links, $L_\mathrm{B}$ and $L_\mathrm{D}$ respectively. 
For a MST move $\Delta L^{\mathrm{MST}}=L_\mathrm{B}^\mathrm{MST}-L_\mathrm{D}^\mathrm{MST}$, while $H^{\mathrm{dMST}}=L_\mathrm{B}^{\mathrm{dMST}}$, 
thus the condition becomes $L_\mathrm{B}^\mathrm{MST}-\beta L_\mathrm{D}^\mathrm{MST} \gtrless L_\mathrm{B}^\mathrm{dMST}$.

\begin{figure}
\centering
\includegraphics{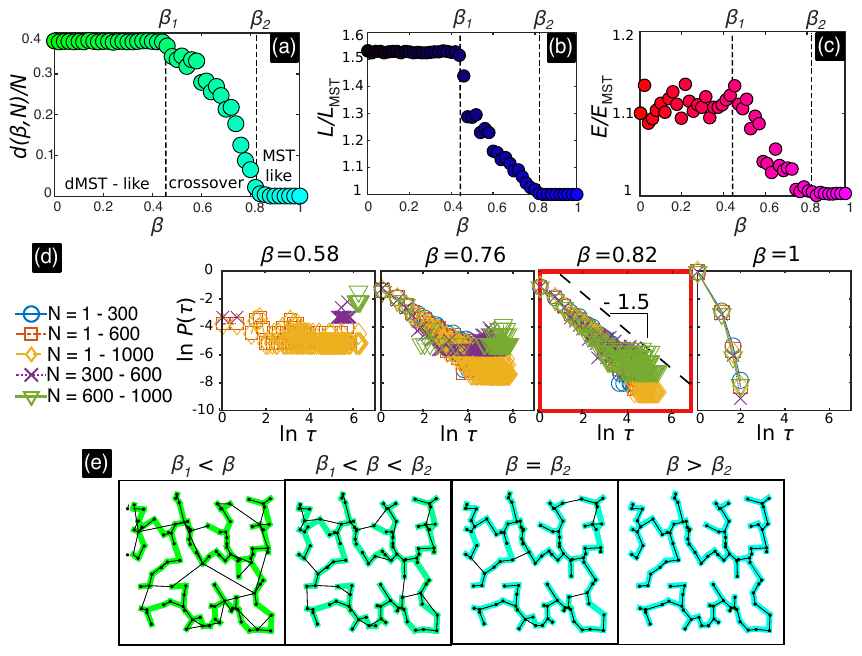}
\caption{(Color online) Structural properties and dynamical evolution of networks grown with different strategies $\beta$. 
The normalized (a) Hamming distance, (b) total length, and (c) efficiency reveal three classes of strategies: MST-like, crossover, and dMST-like, separated by two transition points $\beta_1$ and $\beta_2$. 
(d) The probability distribution of the waiting time between two consecutive MST moves $P(\tau)$ for different windows of network size. In $\beta_2$, $P(\tau)$ is a power-law of exponent $\approx 1.5$ (highlighted box).
(e) Realizations of the model (black thin line) for values of $\beta$ in the three classes and for $\beta=\beta_2$ on the same 50 nodes sequence (black dots), superposed to the corresponding MST (light bold line).
}
\label{fig:2}
\end{figure}

\textit{Results ---}
For each value of $\beta$ from 0 to 1 (by steps of 0.02), we numerically grow 70 networks 
up to $N_\mathrm{f}=1000$ nodes by the rules of the model.
Results are averaged over these 70 networks [Supplemental Material (SM)].
For each network, we measure the normalized Hamming distance $d(\beta,N)/N$, defined as
the number of links that one has to create (and destroy) in order to turn
the network into the MST spanning the same set of nodes, divided by the size of the network $N$ [SM].
This quantity identifies three classes of strategies separated by two transition points, $\beta_1 \approx 0.45$ and $\beta_2\approx 0.82$ [Fig.~\ref{fig:2}(a))].
``MST-like'' strategies ($\beta>\beta_2$) grow networks at very small Hamming distance from the corresponding MST ($\beta = 1$).
``dMST-like'' strategies ($\beta<\beta_1$) grow networks similar to the one grown by iterating dMST moves only ($\beta = 0$).
``Crossover'' strategies ($\beta_1 < \beta < \beta_2$) smoothly interpolate from one extreme to the other.
The phase boundaries and the value of $d(\beta,N)/N$ 
do not depend sensibly on network size after $N \approx 200$
[SM].

The existence of three classes is further confirmed by looking at the total length $L$ and efficiency $E$ of the same networks, normalized by the corresponding MST values and as a function of $\beta$ [Fig.~\ref{fig:2}(b) and (c) and SM].
Efficiency quantifies how quickly information and resources are exchanged over a 
transport network~\cite{Latora:2001kq, cook2014efficiency}, and is often regarded as one of the main design goals in planning and building these networks \cite{GastnerNewman:2006, bebber2007biological}.
It is known that maximizing efficiency competes with minimizing total length~\cite{Latty2014}.
Interestingly, our approach reveals that balancing building and maintenance costs entails a tradeoff between total length and efficiency [Fig.~\ref{fig:2}(b) and (c)],
suggesting that the bias towards efficient transport observed in real networks
may emerge under more general conditions, via optimization of a function of length alone.

To better characterize the classes of strategies observed, we introduce the \emph{waiting time} $\tau$, 
defined as the number of steps
from $N=1$ to the first MST move, and then between two consecutive MST moves.
Due to non-stationarity of the process, the probability distribution function $P(\tau)$ of the waiting time depends not only on $\beta$, but also on the size of the network [Fig.~\ref{fig:2}(d) and SM].
Before $\beta_1$, $P(\tau)$ is not defined, as the typical waiting times are larger than those attained by our simulations ($N_f=1000$). Accordingly, the total length is never minimized through a MST move, and networks in this regime share only a few links with the corresponding MST, typically the shortest ones [Fig.~\ref{fig:2}(e)].
A mean field estimate of $\beta_1$ can be obtained by using the condition for choosing a MST move $L_\mathrm{B}^\mathrm{MST}-\beta L_\mathrm{D}^\mathrm{MST} < L_\mathrm{B}^\mathrm{dMST}$, and assuming that, when $\beta$ is close to $\beta_1$ from above, a MST move destroys and re-builds nearly all the network's links [Fig.~\ref{fig:2}(e)]. The typical length of a link in a MST of $N$ nodes can be estimated as 
the average nearest-neighbor distance among $N$ random points, i.e.
$\sqrt{1/c N}$, where $c$ is some constant. 
Thus, $L_B^{MST} \sim N \sqrt{1/c N}=\sqrt{N/c}$ 
and $L_\mathrm{B}^\mathrm{dMST} \sim \sqrt{1/cN}$,
while $L_D^{MST} \sim \sum_{n=1}^N \sqrt{1/c n} \sim 2\sqrt{N/c}$. The left-hand side of the growth condition becomes $(1-2\beta)\sqrt{N/c}$. Since $L_\mathrm{B}^\mathrm{dMST}$ goes to zero for large $N$, the condition for at least one MST move to occur in this limit becomes $\beta>1/2$, which is not far from the observed $\beta_1 \approx 0.45$.

The optimization condition also suggests that, at the onset of the crossover regime, the occurrence of a MST event is tied to the destruction of long links to build short ones.
Accordingly, $P(\tau)$ shows MST events are rare and happen typically at large network size [Fig.~\ref{fig:2}(d)], where the difference between long links, built in the initial steps, and links that would be built in the MST is large.
At increasing $\beta$, the probability of shorter waiting times increases for small network size, and MST events occur more likely.
In the MST-like phase ($\beta\gtrsim\beta_2$), the growth condition is satisfied often, and $P(\tau)$ decays 
sub-polinomially (exponentially for $\beta=1$) at all network sizes.
Remarkably, the dynamics displays long-range memory
at the transition to the minimum-length phase $\beta_2$. 
Here $P(\tau)$ is a power law of exponent $\approx -1.5$ at all sizes,
and the waiting time $\tau$ has no typical scale (except the cut-off)
contrary to the MST-like and dMST-like phases.
As a consequence, the occurrence of a MST event
is highly unpredictable at $\beta_2$, where waiting times are scale free.
(For further discussion see the SM.)

\begin{figure}
\centering
\includegraphics{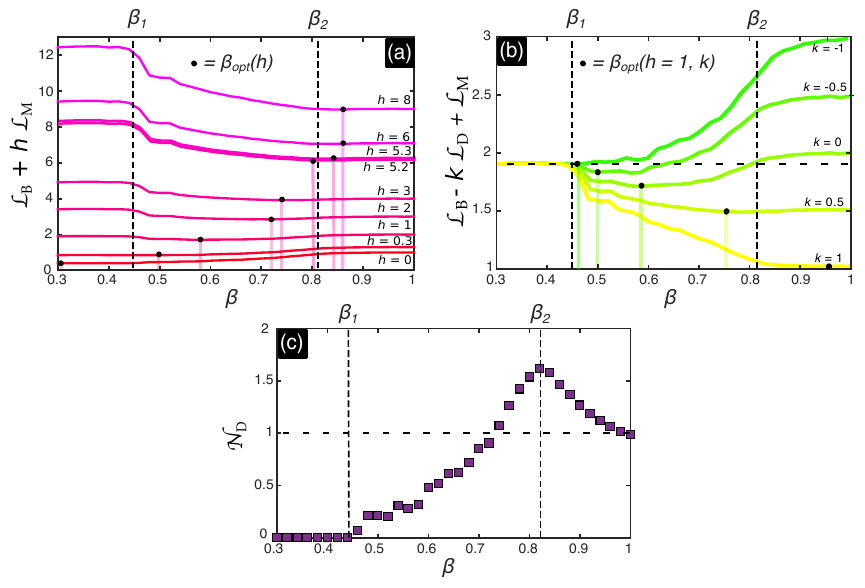}
\caption{(Color online) 
The integrated (up to $N_f=1000$) cost landscapes (solid lines) for
(a) building ($\mathcal{L}_\mathrm{B}$) and maintaining ($\mathcal{L}_\mathrm{M}$), at different values of the ratio $h$ between their unit costs;
(b) building, maintaining, and destroying ($\mathcal{L}_\mathrm{D}$) for $h = 1$. $k \in [-1, 1]$ is the unit cost of destroying material (if $k <0$) or the advantage of recycling (if $k > 0$).
The minimum of each cost landscape (dots) is the optimal strategy $\beta_{opt}$ for the given value of $h$ and $k$. 
(c) The integrated number of links that are destroyed and rebuilt ($\mathcal{N}_\mathrm{D}$) is maximum in $\beta_2$, revealing high non-extensive costs. Strategies above the horizontal dashed line destroy more links than the pure strategy $\beta = 1$.
}
\label{fig:3}
\end{figure}

We analyze the performances of different growth strategies in terms 
of their long-term total cost
by means of three time-integrated quantities $\mathcal{L}_\mathrm{B},\mathcal{L}_\mathrm{D},\mathcal{L}_\mathrm{M}$, defined as
\begin{equation}
\label{eq:costs}
\mathcal{L}_\mathrm{*}(N_f,\beta) = \sum_{N=1}^{N_f} L_\mathrm{*}(N,\beta) / \sum_{N=1}^{N_f} L_\mathrm{*}(N,\beta=1).
\end{equation}
These quantities measure how much length was built ($\mathrm{*} =$ B), destroyed ($\mathrm{*} =$ D), or maintained ($\mathrm{*} =$ M) up to $N_f = 1000$ by each strategy $\beta$. 
$L_\mathrm{B}(N,\beta)$ and $L_\mathrm{D}(N,\beta)$ 
are the instantaneous lengths built and destroyed between step $N-1$ and step $N$,
and $L_\mathrm{M}(N,\beta)$ is the total length of the network at size $N$. 
All the measures are normalized by the values they take in a pure MST dynamics (i.e., at $\beta=1$) with the same realization of the point process.

In the simple scenario where the costs of maintenance and building per unit length have ratio $h$, the final cost of a network is given by $\mathcal{L}_\mathrm{B} + h \, \mathcal{L}_\mathrm{M}$.
Plotting this total cost against $\beta$ produces a cost landscape for each value of $h$ [Fig.~\ref{fig:3}(a)]. 
Each cost landscape has an absolute minimum, which identifies the optimal strategy $\beta_{opt}(h)$ for the given ratio $h$.
Interestingly, crossover strategies turn out to be optimal for a wide range of values of the ratio $h$ ($0.3 \lesssim h \lesssim 5.2$). More complicated cost scenarios can be analyzed.
For example, one may consider that
building costs during a MST move may be reduced
by recycling the material obtained from the destruction of existing links.
On the contrary, when recycling is not possible, disposing of the destroyed material may bear additional costs.
Such scenarios can be described by adding the time-integrated destroyed length to the total cost: 
$\mathcal{L}_\mathrm{B} +h\,\mathcal{L}_\mathrm{M} - k \, \mathcal{L}_\mathrm{D}$. 
The coefficient $k \in [-1, 1]$ is the fraction of destroyed material that can be recycled (if $k>0$) or bearing additional disposal costs (if $k<0$).
Also in this scenario, crossover strategies play an important role in minimizing the total efforts for construction and maintenance of transport networks [Fig.~\ref{fig:3}(b), particular case of $h=1$], realizing nontrivial tradeoffs between the competing costs.
The optimal strategy is in the crossover regime even when destroying is as expensive as building and maintaining ($k=-1$), while MST-like strategies are optimal only when recycling strongly lowers the total cost.

All costs considered above are extensive in the length of the transport channels involved.
However, length-independent costs may be present in empirical situations, for instance associated to setting up the
sites for building and dismantling connections.
These ``fixed'' costs depend on the number of links modified at each step, regardless of their length.
We quantify these non-extensive costs via the total number of links that were destroyed (and rebuilt)
$\mathcal{N}_\mathrm{D}(N,\beta) = \sum_{n=1}^{N} N_\mathrm{D}(n,\beta) / \sum_{n=1}^{N} N_\mathrm{D}(n,\beta=1)$. 
$N_\mathrm{D}$ is the number of links destroyed at each step, and the sum is normalized by the corresponding MST value,
as in (\ref{eq:costs}).
Interestingly, $\beta_2$ 
is the strategy requiring the largest number of link deletions,
and is therefore a point of strong non-optimality in terms of fixed costs [Fig.~\ref{fig:3}(c)].
Crossover strategies with $\beta \gtrsim 0.75$ and MST-like strategies require to destroy (and thus to re-build) more links 
than in the pure MST strategy [Fig.~\ref{fig:3}(c) horizontal dashed line].

\begin{figure}
\centering
\includegraphics{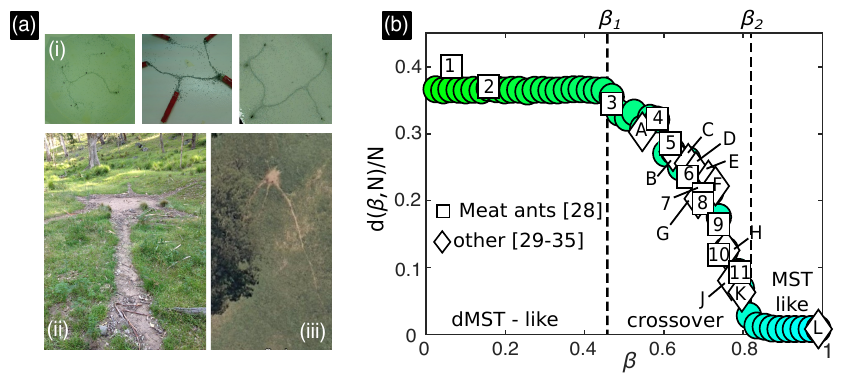}
\caption{(Color online) (a) Cost-optimized ant networks: (i) Argentine ants find the shortest path to connect their nests. Credit: Tanya Latty. (ii) Meat ant's nest with departing trails. Credit: Nathan Brown. (iii) Part of a meat ant's transport network from Google Earth. (b) Comparing the Hamming distance of ant transport networks (meat ants \cite{Wilg2007colonystructure} are squares, other species \cite{heller2008nest, holt1990observations,andersenpatel,mciver1991dispersed, boudjema2006analysis, Ellis01092014, pfeiffer2000contributions} are diamonds) with our model suggests crossover strategies are relevant for different ant species.
}
\label{fig:4}
\end{figure}

\textit{Discussion ---}
Albeit simple, our interpolating model presents a rich behavior, providing a general framework to understand 
the competing nature of construction and maintenance costs.
In doing so, it addresses the interplay of central planning and local growth characterizing the growth of many man-made transport networks, 
offering insights in the long-term outcome of different short-term construction strategies. 
Unexpectedly, intermediate growth strategies are optimal in many cost scenarios, as they minimise the long-term total costs entailed by the infrastructure.
Moreover, we showed balancing competing costs is a minimal sufficient ingredient for the emergence of the 
tradeoff between the network's total length and its transport efficiency, 
which is usually explained by more system-specific principles.
Finally, the model displays a transition point with diverging characteristic time, 
similarly to the phenomenon of critical slowing down close to
phase transitions in statistical mechanics, which maximizes the long-term number of links rewired.

A key premise in the formulation of the model is that the position of new nodes
is not known beforehand.
If the time scale of the arrival of new nodes is much larger than that of the transport processes
on the network itself, then each new node needs to be connected before
the position of successive nodes can be taken into account.
A possible example in human systems is
the evolution of bus routes
\cite{dodson2011principles, sekhar2004approach}.
New areas can be quickly connected by adding further segments to bus lines stopping nearby (dMST move). 
However, if the whole network becomes suboptimal in terms of running costs,
it may become necessary to re-design it globally (MST move).
Our model suggests there may be an optimal re-organization frequency that minimizes the total costs of bus route networks.

In nature, a striking example of cost-constrained transport networks are the trails built by polydomous ant colonies to connect spatially separated nests~\cite{perna2012}.
Under laboratory conditions, the Argentine ant \emph{Linepithema humile} builds globally optimized transport networks that resemble MST or even Steiner trees (minimum spanning trees where the set of nodes is allowed to be enlarged)~\cite{latty2011structure} [Fig.~\ref{fig:4}(a), top].
Conversely, the Australian meat ant \emph{Iridomyrmex purpureus} tends to link 
each newly built nest to the closest one in the colony~\cite{Bottinelli2015},
as in the dMST move in our model [Fig.~\ref{fig:4}(a), bottom].
For these ants, it has been observed that, during colony growth, suboptimal connections can be progressively substituted with shorter ones and eventually abandoned \cite{greaves1974},
realising a dynamics similar to the one implemented by our model
(although on a time scale comparable to that of node addition). 
Building on these observations, we used our model as a framework to 
quantify the trade-off between building and maintenance
costs experienced by ants.
For 30 published networks constructed by different ant species 
({\it Linepithema humile} \cite{heller2008nest}, {\it Iridomyrmex purpureus} \cite{holt1990observations,andersenpatel,mciver1991dispersed,Wilg2007colonystructure}, {\it Formica lugubris} \cite{boudjema2006analysis, Ellis01092014} and {\it Camponotus gigas} \cite{pfeiffer2000contributions}, see SM for detailed description of the datasets and methods), 
we measured the normalized Hamming distance from the MST built on the same set of nodes, and assigned a strategy $\beta$ to each network
by comparison with the model prediction for $d(\beta,N)/N$ [Fig.~\ref{fig:4}(b)]. 
In all but one of the analyzed trail networks, the rescaled distance $d(\beta,N)/N$ from the corresponding MST
departs from $0$ at most as much as the typical dMSTs ($d(\beta,N)/N \approx 0.38$).
Interestingly, for 27 colonies out of 30, the estimated strategy $\beta$ is in the crossover regime (as a consequence of the distribution of their rescaled distances).
This suggests both maintenance and building costs are relevant
in the growth of ant networks, and that different species may share
common underlying building principles and optimization strategies. 
Moreover, the balance between efficiency and total length characterizing most empirical ant networks~\cite{Latty2014}, may be explained by such cost-constrained strategies alone.
Alternating between local and global interventions on the network 
may thus confer evolutionary advantages,
and should be taken into account when analysing transport networks.

\section*{Acknowledgements}
We would like to thank D.J.T.~Sumpter, J.L.~Silverberg, J.L.~Denebourg, and M.~Cosentino Lagomarsino 
for constructive feedback and stimulating discussions. 
A.B. acknowledges funding from the Centre
for Interdisciplinary Mathematics (CIM).

\bibliographystyle{unsrt}
\bibliography{spatial_networks}

\begin{thebibliography}{10}

\bibitem{katifori2016}
H.~Ronellenfitsch and E.~Katifori.
\newblock Global optimization, local adaptation, and the role of growth in
  distribution networks.
\newblock {\em Phys. Rev. Lett.}, 117:138301, Sep 2016.

\bibitem{tero2010rules}
A.~Tero, S.~Takagi, T.~Saigusa, K.~Ito, D.~P. Bebber, M.~D. Fricker, K.~Yumiki,
  R.~Kobayashi, and T.~Nakagaki.
\newblock Rules for biologically inspired adaptive network design.
\newblock {\em Science}, 327(5964):439--442, 2010.

\bibitem{perna2012}
A.~{Perna}, B.~{Granovskiy}, S.~{Garnier}, S.~C. {Nicolis}, M.~{Lab{\'e}dan},
  G.~{Theraulaz}, V.~{Fourcassi{\'e}}, and D.~J.~T. {Sumpter}.
\newblock {Individual Rules for Trail Pattern Formation in Argentine Ants
  (Linepithema humile)}.
\newblock {\em PLoS Computational Biology}, 8:2592, July 2012.

\bibitem{gastner2006shape}
M.~T. Gastner and M.~E.J. Newman.
\newblock Shape and efficiency in spatial distribution networks.
\newblock {\em Journal of Statistical Mechanics: Theory and Experiment},
  2006(01):P01015, 2006.

\bibitem{buhl2006topological}
J.~Buhl, J.~Gautrais, N.~Reeves, R.V. Sol{\'e}, S.~Valverde, P.~Kuntz, and
  G.~Theraulaz.
\newblock Topological patterns in street networks of self-organized urban
  settlements.
\newblock {\em The European Physical Journal B-Condensed Matter and Complex
  Systems}, 49(4):513--522, 2006.

\bibitem{Gorman2004}
S.~P. Gorman and R.~Kulkarni.
\newblock Spatial small worlds: new geographic patterns for an information
  economy.
\newblock {\em Environment and Planning B: Planning and Design},
  31(2):273--296, 2004.

\bibitem{vessels1}
G.~Serini, D.~Ambrosi, E.~Giraudo, A.~Gamba, L.~Preziosi, and F.~Bussolino.
\newblock Modeling the early stages of vascular network assembly.
\newblock {\em The EMBO Journal}, 22(8):1771--1779, 04 2003.

\bibitem{bebber2007biological}
D.~P. Bebber, J.~Hynes, P.~R. Darrah, L.~Boddy, and M.~D. Fricker.
\newblock Biological solutions to transport network design.
\newblock {\em Proceedings of the Royal Society B: Biological Sciences},
  274(1623):2307--2315, 2007.

\bibitem{buhl2009shape}
J.~Buhl, K.~Hicks, E.~R. Miller, S.~Persey, O.~Alinvi, and D.~J.T. Sumpter.
\newblock Shape and efficiency of wood ant foraging networks.
\newblock {\em Behavioral Ecology and Sociobiology}, 63(3):451--460, 2009.

\bibitem{xie2009modeling}
F.~Xie and D.~Levinson.
\newblock Modeling the growth of transportation networks: a comprehensive
  review.
\newblock {\em Networks and Spatial Economics}, 9(3):291--307, 2009.

\bibitem{Barthelemy2011}
M.~Barth{\'e}lemy.
\newblock Spatial networks.
\newblock {\em Physics Reports}, 499(1--3):1 -- 101, 2011.

\bibitem{GastnerNewman:2006}
M.T. Gastner and M.E.J. Newman.
\newblock Optimal design of spatial distribution networks.
\newblock {\em Phys. Rev. E}, 74:016117, 2006.

\bibitem{BanavarMaritan:1999}
J.R. Banavar, A.~Maritan, and A.~Rinaldo.
\newblock Size and form in efficient transportation networks.
\newblock {\em Nature}, 399:130, 1999.

\bibitem{MaritanColaiori:1996}
A.~Maritan, F.~Colaiori, A.~Flammini, M.~Cieplak, and J.R. Banavar.
\newblock Universality classes of optimal networks.
\newblock {\em Science}, 272:984--986, 1996.

\bibitem{Guimera:2002}
R.~Guimer{\`a}, A.~Diaz-Guilera, F.~Vega-Redondo, A.~Cabrales, and A.~Arenas.
\newblock Optimal network topologies for local search with congestion.
\newblock {\em Phys. Rev. Lett.}, 89:248701, 2002.

\bibitem{Colizza:2004}
V.~Colizza, J.R. Banavar, A.~Maritan, and A.~Rinaldo.
\newblock Network structures from selection principles.
\newblock {\em Phys. Rev. Lett.}, 92:198701, 2004.

\bibitem{Bottinelli2015}
A.~Bottinelli, E.~van Wilgenburg, D.~J.~T. Sumpter, and T.~Latty.
\newblock Local cost minimization in ant transport networks: from small-scale
  data to large-scale trade-offs.
\newblock {\em Journal of The Royal Society Interface}, 12(112), 2015.

\bibitem{FerreriCancho:2003}
R.~Ferrer~i Cancho and R.V. Sol{\'e}.
\newblock Optimization in complex networks.
\newblock In {\em Statistical Mechanics of Complex Networks}, volume 625 of
  {\em Lecture Notes in Physics}, pages 114--125. Springer, 2003.

\bibitem{Brede:2010}
M.~Brede.
\newblock Coordinated and uncoordinated optimization of networks.
\newblock {\em Phys. Rev. E}, 81:066104, 2010.

\bibitem{Brede:2010b}
M.~Brede.
\newblock Optimal synchronization in space.
\newblock {\em Phys. Rev. E}, 81:025202(R), 2010.

\bibitem{BarthelemyFlammini:2006}
M.~Barth{\'e}lemy and A.~Flammini.
\newblock Optimal traffic networks.
\newblock {\em J. Stat. Mech.}, (07):L07002, 2006.

\bibitem{fabrikant2002heuristically}
A.~Fabrikant, E.~Koutsoupias, and C.~H. Papadimitriou.
\newblock Heuristically optimized trade-offs: A new paradigm for power laws in
  the internet.
\newblock In {\em Automata, languages and programming}, volume 2380 of {\em
  Lecture Notes in Computer Science}, pages 110--122. Springer, 2002.

\bibitem{louf2013emergence}
R.~Louf, P.~Jensen, and M.~Barthelemy.
\newblock Emergence of hierarchy in cost-driven growth of spatial networks.
\newblock {\em Proceedings of the National Academy of Sciences},
  110(22):8824--8829, 2013.

\bibitem{1980transport}
P.~E. O'Sullivan.
\newblock {\em Transport policy: geographic, economic, and planning aspects}.
\newblock Rowman and Littlefield, 1980.

\bibitem{taplin2005cost}
J.~H.E. Taplin, M.~Qui, V.~K. Salim, and R.~Han.
\newblock {\em Cost-benefit analysis and evolutionary computing: optimal
  scheduling of interactive road projects}.
\newblock Edward Elgar Publishing, 2005.

\bibitem{louzada2012generating}
V.~H.P. Louzada, F.~Daolio, H.~J. Herrmann, and M.~Tomassini.
\newblock Generating robust and efficient networks under targeted attacks.
\newblock {\em arXiv preprint arXiv:1207.1291}, 2012.

\bibitem{Barabasi2000}
R.~Albert, H.~Jeong, and A.~Barabasi.
\newblock Error and attack tolerance of complex networks.
\newblock {\em Nature}, 406(6794):378--382, 07 2000.

\bibitem{Latora:2001kq}
V.~Latora.
\newblock Efficient behavior of small-world networks.
\newblock {\em Physical Review Letters}, 87(19), 2001.

\bibitem{cook2014efficiency}
Z.~Cook, D.~W. Franks, and E.~J.H. Robinson.
\newblock Efficiency and robustness of ant colony transportation networks.
\newblock {\em Behavioral Ecology and Sociobiology}, 68(3):509--517, 2014.

\bibitem{Latty2014}
G.~Cabanes, E.~van Wilgenburg, M.~Beekman, and T.~Latty.
\newblock Ants build transportation networks that optimize cost and efficiency
  at the expense of robustness.
\newblock {\em Behavioral Ecology}, page aru175, 2014.

\bibitem{Wilg2007colonystructure}
E.~Van~Wilgenburg and M.A. Elgar.
\newblock Colony structure and spatial distribution of food resources in the
  polydomous meat ant iridomyrmex purpureus.
\newblock {\em Insectes sociaux}, 54(1):5--10, 2007.

\bibitem{heller2008nest}
N.E. Heller, K.K. Ingram, and D.M. Gordon.
\newblock Nest connectivity and colony structure in unicolonial argentine ants.
\newblock {\em Insectes Sociaux}, 55(4):397--403, 2008.

\bibitem{holt1990observations}
J.A. Holt.
\newblock Observations on the relationships between meat ants and termites in
  tropical australia.
\newblock {\em Journal of tropical ecology}, 6(03):379--382, 1990.

\bibitem{andersenpatel}
A.~N. Andersen and A.~D. Patel.
\newblock Meat ants as dominant members of australian ant communities: an
  experimental test of their influence on the foraging success and forager
  abundance of other species.
\newblock {\em Oecologia}, 98(1).

\bibitem{mciver1991dispersed}
J.D. McIver.
\newblock Dispersed central place foraging in australian meat ants.
\newblock {\em Insectes Sociaux}, 38(2):129--137, 1991.

\bibitem{boudjema2006analysis}
G.~Boudjema, G.~Lemp{\'e}ri{\`e}re, M.~Deschamps-Cottin, and D.~G. Molland.
\newblock Analysis and nonlinear modeling of the mound-building ant formica
  lugubris spatial multi-scale dynamic in a larch-tree stand of the southern
  french alps.
\newblock {\em Ecological modelling}, 190(1), 2006.

\bibitem{Ellis01092014}
S.~Ellis, D.~W. Franks, and E.~J.H. Robinson.
\newblock Resource redistribution in polydomous ant nest networks: local or
  global?
\newblock {\em Behavioral Ecology}, 25(5), 2014.

\bibitem{pfeiffer2000contributions}
M.~Pfeiffer and K.E. Linsenmair.
\newblock Contributions to the life history of the malaysian giant ant
  camponotus gigas (hymenoptera, formicidae).
\newblock {\em Insectes Sociaux}, 47(2):123--132, 2000.

\bibitem{dodson2011principles}
J.~Dodson, P.~Mees, J.~Stone, and M.~Burke.
\newblock The principles of public transport network planning: A review of the
  emerging literature with select examples.
\newblock {\em Issues paper}, 15, 2011.

\bibitem{sekhar2004approach}
S.V.C. Sekhar, W.~L. Yue, and M.A.P. Taylor.
\newblock An approach to transit path design using gis.
\newblock {\em International Journal of Urban Sciences}, 8(1):28--39, 2004.

\bibitem{latty2011structure}
T.~Latty, K.~Ramsch, K.~Ito, T.~Nakagaki, D.~J.T. Sumpter, M.~Middendorf, and
  M.~Beekman.
\newblock Structure and formation of ant transportation networks.
\newblock {\em Journal of The Royal Society Interface}, 8(62):1298--1306, 2011.

\bibitem{greaves1974}
T.~Greaves and R.~D. Hughes.
\newblock The population biology of the meat ant.
\newblock {\em Australian Journal of Entomology}, 13(4):329--351, 1974.

\end{thebibliography}

\end{document}